\documentclass[a4paper,fleqn]{cas-dc}

\usepackage[numbers]{natbib}
\usepackage{cite}
\def\tsc#1{\csdef{#1}{\textsc{\lowercase{#1}}\xspace}}
\tsc{WGM}
\tsc{QE}
\tsc{EP}
\tsc{PMS}
\tsc{BEC}
\tsc{DE}

\begin{document}
\let\WriteBookmarks\relax
\def\floatpagepagefraction{1}
\def\textpagefraction{.001}

\shorttitle{Warm inflation in non-standard backgrounds}
\shortauthors{Oem Trivedi}

\title [mode = title]{The exact solution approach to warm inflation}

\author[1]{Oem Trivedi}[]
\cormark[1]

\ead{oem.t@ahduni.edu.in}

\affiliation[1]{organization={International Centre for Space and Cosmology, School of Arts and Sciences, Ahmedabad University},
    addressline={Navrangpura}, 
    city={Ahmedabad},
    postcode={380009},
    country={India}}

\cortext[cor1]{Corresponding author}

\begin{abstract}
The theory of cosmic inflation has received a great amount of deserved attention in recent years due to it's stunning predictions about the early universe. Alongside the usual cold inflation paradigm, warm inflation has garnered a huge amount of interest in modern inflationary studies. It's peculiar features and specifically different predictions from cold inflation have led to a substantial amount of literature about it. Various modified cosmological scenarios have also been studied in the warm inflationary regime. In this work, we introduce the exact solution approach for warm inflation. This approach allows one to directly study warm inflationary regime in a variety of modified cosmological scenarios. We begin by outlining our method and show that it generalizes the modified Friedmann approach of Del Campo , and reduces to the well known Hamilton-Jacobi formalism for inflation in particular limits. We also find the perturbation spectra for cosmological and tensor perturbations in the early universe, and then apply our method to study warm inflation in a Tsallis entropy modified Friedmann universe. We end our paper with some concluding remarks on the domain of applicability of our work.
\end{abstract}

\begin{keywords}
early universe cosmology \sep warm inflation \sep modified gravity theories
\end{keywords}

\maketitle

\section{Introduction}
	
	Some of the most captivating problems in modern cosmology concern the workings of the very early Universe. A lot of the physics of how our Universe was initially is not very well understood and this was evident ever since fine tuning problems were discovered in the traditional big bang cosmology model itself. The most prominent of these problems were the flatness, horizon and monopole problems, which were very promisingly solved by the introduction of Cosmic inflation \citep{guth1981inflationary,albrecht1982reheating,linde1983chaotic,sato1981first,Clifton:2011jh,Nojiri:2010wj,Capozziello:2011et,Odintsov:2023weg,Nojiri:2017ncd,Sotiriou:2008rp}. Cosmic inflation is the idea that the very early Universe went through a period of rapid accelerated expansion. The dynamics of the traditional inflationary models generally consider a real scalar field, the "inflaton" field, as the predominant contributor to the universal energy density at that time. While there are other inflationary models which consider more exotic scenarios \citep{golovnev2008vector, scialom1996inflation,linde1994hybrid}, there is still a very high amount of evident interest in single real scalar field driven inflationary regimes \citep{baumann2009tasi,linde1995lectures,brandenberger2001frontiers,linde2005particle}. \\
	\\
	The conventional models of inflation by a single scalar field consist of a period of rapid accelerated expansion. The sub horizon scale quantum fluctuations during this period of expansion, lead to cosmological perturbations which give us a spectrum of scalar and tensor perturbations ( vector perturbations are usually ignored). This generates the large scale structure of our Universe which we see today. Also , after the period of expansion sets off a period of " reheating" \citep{albrecht1982reheating} , where the energy of the inflaton field decays to form radiation energy densities. These models do not take into account the dissipation of inflaton energy to form radiation while the inflationary period itself is going on and hence, these are known popularly now as " cold inflation" scenarios. On the other hand, the class of inflationary regimes which do take into account this dissipation effect are known as " warm inflation" regimes \citep{berera1995warm,berera2009warm,rangarajan2018current,berera2004warm}. Warm inflation includes inflaton interactions with other fields throughout the inflationary epoch instead of confining such interactions to a distinct reheating era. Consequently, these inflationary models do not need a separate period of reheating after inflation to make the universe radiation dominated, and hence this provides a solution to the graceful exit problem of inflation \citep{berera1997interpolating}.
	\\
	\\
	Recently a lot of interest has been weighed towards cosmological models which modify the structure of the usual Friedmann Equation \begin{equation}
 		H^{2} = \frac{8 \pi}{3 m_{p}^{2}} \rho 
	\end{equation}  where we have used the units $ c = \hbar = 1 $ and $ m_{p}^{2} $ is the usual reduced Planck mass. $ \rho $ is the energy density in the Universe, which in the context of usual Inflationary Models refers to the energy density of the inflaton field. Various different cosmological scenarios bring about a change in this usual Friedmann equation . Superstring and M-Theory bring the possibility of considering our universe as a domain wall embedded in a higher dimensional space. In this scenario the standard model of particle is confined to the brane, while gravitation propagate into the bulk spacetime. The effect of extra dimensions induces a well known change in the Friedmann equation \citep{shiromizu2000einstein} ,
	\begin{equation}
	 H^{2} = \frac{8 \pi}{3 m_{p}^{2}} \rho (1 + \frac{\rho}{2 \lambda})
	\end{equation} where $\lambda$ is a measure of the brane tension. 
	\\
	\\
	It is also possible to consider quantum corrected area entropy relations and study universal evolution in such scenarios. For example, \citep{cai2008corrected} considered the quantum corrected Entropy \begin{equation}
	S = m_{p}^{2} \frac{A}{4} - \tilde{\alpha} \ln (m_{p}^{2} \frac{A}{4}) 
	\end{equation} where A is the area of the apparent horizon and $ \tilde{\alpha} $ is dimensionless positive constant determined by conformal anolmaly of the fields.This
	conformal anomaly is interpreted as a quantum correction to the entropy of the apparent horizon. The resulting Friedmann equation is of the form \begin{equation}
	H^{2} - \beta H^{4} = \frac{8 \pi}{3 m_{p}^{2}} \rho 
	\end{equation} , where $\beta$ is some positive constant in the units of inverse mass squared. This type of Friedmann equation can also be arrived at by using a Chern-Simons modification of gravity \citep{gomez2011standard} . Finally, there has been a surge in interest recently in the form of Black hole Entropy due to Tsallis and Cirto \citep{tsallis2013black} , where they argued that argued that the microscopic mathematical expression of the thermodynamical
	entropy of a black hole does not obey the area law and
	can be modified as \begin{equation}
	S = \gamma A^{\kappa}
	\end{equation} , where $\gamma$ is an unknown constant , A is the Black Hole Horizon area and $\kappa$ is a real parameter which quantifies the degree of nonextensivity, known as the Tsallis parameter. Considering the apparent horizon entropy of an FLRW type universe to be the Tsallis type, \citep{sheykhi2018modified} was able to derive a Friedmann equation for the universe while also taking into account and modifying Padmanabhan's emergent gravity proposal \citep{padmanabhan2012emergence}. The concerned Friedmann equation is \citep{sheykhi2018modified} , \begin{equation} \label{6}
	(H^{2})^{2 - \kappa} = \frac{8 \pi }{3 m_{p}^{2}} \rho 
	\end{equation}
	Inflation has been studied in a lot of the popular braneworld scenarios , both in the traditional supercooled regime \citep{choudhury2003hybrid,hawkins2001inflation,kallosh2003p} and in the warm inflation regime \citep{kamali2020warm,cid2007warm,del2007warm,del2012approach}. \citep{del2012approach} also studied Chern-Simons type modified gravity of \citep{gomez2011standard} in the cold inflation regime. A crucial point , as noted in \citep{del2012approach}, which emerges from the studies of inflationary studies in different cosmological scenarios is that the concerned Friedmann equation of investigation can be often taken of the form \begin{equation}\label{7}
	F(H) = \frac{8 \pi}{3 m_{p}^{2}} \rho 
	\end{equation} , where $ F(H) $ is some function of the Hubble parameter H. A very useful approach to studying different cold inflation scenarios was discussed in \citep{del2012approach}. We would now like to study different cosmological scenarios using \eqref{7} as the primary equation of investigation. We hence structure our paper as follows. In section 2, we outline the exact solution approach for Warm Inflation and describe the dynamical aspects of the method while in Section 3 we discuss Perturbation spectra and derive crucial observational quantities like the scalar and tensor spectral index in concern to the approach. In Section 4 we apply the method to study Warm Inflation in the high dissipative regime in a Tsallis entropy type Universe of \citep{sheykhi2018modified} . Then finally, we summarize our results in section 5 in the Concluding remarks of the paper.
	\\
	\section{The exact solution method}
	In warm inflationary approach, there is consideration given to both the inflaton field and the radiation being created due to it's dissipation during Inflation itself. Hence, the usual Friedmann equation of warm inflation is \citep{berera1995warm} \begin{equation} \label{9}
	H^{2} = \frac{8 \pi}{3 m_{p}^{2}} (\rho_{\phi} + \rho_{r}) 
	\end{equation} where the energy density $\rho$ consists of both inflaton and radiation energy densities, $ \rho_{\phi} $ and $ \rho_{r} $ respectively. But during inflation , $ \rho_{\phi} >> \rho_{r} $ \citep{berera1995warm}, hence \eqref{9} becomes \begin{equation}
	H^{2} = \frac{8 \pi}{3 m_{p}^{2}} (\rho_{\phi})
	\end{equation} , where $ \rho_{\phi} $ is given by , \begin{equation}
	\rho_{\phi} = \frac{\dot{\phi}^{2}}{2} + V(\phi)
	\end{equation} with $ V(\phi) $ being the potential under which the inflaton field is. The inflaton field equation , the inflaton energy density and radiation energy density evolution equation is given by, \begin{equation} \label{12}
	\ddot{\phi} + (3H + \Gamma) \dot{\phi} + V^{\prime} (\phi) = 0
	\end{equation} \begin{equation}
	\dot{\rho_{\phi}} + 3H(\rho_{\phi} + p_{\phi}) = - \Gamma \dot{\phi}^{2}
	\end{equation} 
	\begin{equation}
	\dot{\rho_{r}} + 4H \rho_{r} = \Gamma \dot{\phi}^{2}
	\end{equation} Here , $\Gamma$ is the dissipation coefficient of inflaton energy. Defining a quantity $ Q = \frac{\Gamma}{3 H} $ , we can rewrite the above equations as \begin{equation}
	\ddot{\phi} + 3H (1 + Q) \dot{\phi} + V^{\prime} (\phi) = 0
	\end{equation} \begin{equation}
		\dot{\rho_{\phi}} + 3H(\rho_{\phi} + p_{\phi}) = - 3HQ \dot{\phi}^{2}
	\end{equation}
	\begin{equation} \label{17}
		\dot{\rho_{r}} + 4H \rho_{r} = 3HQ \dot{\phi}^{2}
	\end{equation}
	During inflation, we can take both $ \dot{\rho_{r}} $ to be approximately zero \citep{berera1995warm} . This gives us \eqref{17} \begin{equation} \label{18}
	\rho_{r} = \frac{3Q}{4} \dot{\phi}^{2}
	\end{equation} 
	This completes a small review of the basics of warm inflation which we will now freely refer to throughout the entirety of paper. Consider now the following general form of the Friedmann equation \begin{equation}\label{19}
	F(H) = \frac{8 \pi}{3 m_{p}^{2}} \rho_{\phi} = \frac{8 \pi}{3 m_{p}^{2}} \left( \frac{\dot{\phi}^{2}}{2} + V(\phi) \right)
	\end{equation} The famous slow roll conditions of cold inflation \citep{baumann2009tasi}, hold well similarly in warm inflation as well \citep{berera1995warm} and so during warm inflation, \begin{equation}
	\dot{\phi}^{2} << V(\phi)
	\end{equation} This gives us the energy density to be $ \rho_{\phi} \approx V(\phi)$ Hence the general Friedmann equation\eqref{19} during Inflation, takes the form \begin{equation} \label{21}
	F(H) \approx \frac{8 \pi}{3 m_{p}^{2}} V(\phi)
	\end{equation} Now, it was shown in \citep{kinney1997hamilton} that treating $\phi$ itself as the explicitly dependent variable for H can lead to a very apt method of studying inflationary models, which is referred to as the " Hamilton-Jacobi approach to inflation". \citep{sayar2017hamilton} studied the Hamilton-Jacobi approach with respect to warm inflation and found that treating the Hubble parameter and the disspiation coefficients in terms of $\phi$ is indeed very helpful for warm inflationary regimes. Hence, now we treat the Hubble parameter ( and hence the scale factor as well) primarily in terms of the field variable $\phi$. So $ H = H (\phi) $. Further the field variable will generally be time dependent, hence the Hubble parameter stays implicitly time dependent. 
	\\
	\\
	 With these considerations, we take the derivative of \eqref{21} with respect to $\phi$ , \begin{equation} \label{22}
	 F_{_{\prime} H} H^{\prime} (\phi) = \frac{8 \pi}{3 m_{p}^{2}} V^{\prime} (\phi)
	 \end{equation} where , following the convention in \citep{del2012approach}, $ F_{_{\prime} H} $ is the partial derivative of F with respect to H  , and the prime in the superscript denotes derivative with respect to $\phi$ . In the slow roll approximation \eqref{12} becomes, \begin{equation}
	 3H(1 + Q) \dot{\phi} = - V^{\prime} (\phi)
 	 \end{equation} Using the expression for $ V^{\prime} (\phi) $ from \eqref{22} we have $ \dot{\phi} $ as \begin{equation} \label{24}
 	 \dot{\phi} = - \frac{m_{p}^{2}}{8 \pi} \frac{F_{_{\prime} H} H^{\prime} }{H (1 + Q)} 
 	 \end{equation} Using \eqref{24} , we can get a different expression for $ \rho_{r} $ by taking into account \eqref{18} , \begin{equation} \label{25}
 	 \rho_{r} = \frac{3Q}{4}  \left(\frac{m_{p}^{2}}{8 \pi} \frac{F_{_{\prime} H} H^{\prime} }{H (1 + Q)} \right)^{2}
 	 \end{equation} Hence, we now have expression for the field velocity given $ F(H) $  ,an ansatz for H in terms of $\phi$ and a form of the dissipation coefficient. As mentioned in \citep{moss2006dissipation,berera2009warm}, dissipation coefficient is popularly used in power laws in terms of both $\phi$ and temperature T. The temperature plays an especially important role for  high dissipative regimes, which is characterized by $$ Q >> 1 \implies \Gamma >> 3H $$ and temperature itself plays a very crucial in the generation of quantum fluctuations during Inflation \citep{berera1997interpolating,berera2009warm}. 
	Using \eqref{25} we can get an expression for $ \rho_{r} $ in terms of $ \rho_{\phi} $ and it's easy to check that \begin{equation} \label{27}
	\rho_{r} = \frac{Q \rho_{\phi} m_{p}^{2}}{32 \pi F} \left( \frac{F_{_{\prime} H} H^{\prime}}{H (1 + Q)} \right)^{2}
	\end{equation}
	We also note that as $ \rho_{r} $ is pure radiation , we can write \begin{equation} \label{28}
	\rho_{r} = \alpha T^{4}
	\end{equation} where T is the radiation temperature and $\alpha$ is the Stefan Boltzmann Constant. Now, we have from \eqref{18} and \eqref{25}, \begin{equation}
	\alpha T^{4} = \frac{3Q}{4}  \left(\frac{m_{p}^{2}}{8 \pi} \frac{F_{_{\prime} H} H^{\prime} }{H (1 + Q)} \right)^{2}
	\end{equation} which gives us \begin{equation} \label{30}
	T =  \left(\frac{3Q}{4 \alpha}\right)^{1/4}  \left(\frac{m_{p}^{2}}{8 \pi} \frac{F_{_{\prime} H} H^{\prime} }{H (1 + Q)} \right)^{1/2}
	\end{equation} 
	We note from the Friedmann equation that \begin{equation}
	\frac{3 m_{p}^{2}}{8 \pi} F - \frac{1}{2} \dot{\phi}^{2} = V(\phi)
	\end{equation} Using the expression for $ \dot{\phi} $ \eqref{24}, we have the potential given as  \begin{equation}
	V(\phi) = \frac{3 m_{p}^{2}}{8 \pi} F - \frac{1}{2} \left(\frac{m_{p}^{2}}{8 \pi} \frac{F_{_{\prime} H} H^{\prime} }{H (1 + Q)} \right)^{2} 
	\end{equation} Also as $$ \rho_{\phi}  = \frac{3 m_{p}^{2}}{8 \pi} F $$ , we can also write the potential in terms of the inflaton energy density \begin{equation}
	V(\phi) = \rho_{\phi} \left( 1 - \frac{ m_{p}^{2}}{24 \pi} \left( \frac{F_{_{\prime} H} H^{\prime}}{H ( 1 + Q )} \right)^{2} \right) 
	\end{equation}
	\\
	The number of e-folds is a very important quantity to measure the amount of Inflation. The number of e-folds is [\citep{baumann2009tasi} ], \begin{equation}
	N = \int_{\phi_{o}}^{\phi_{e} } \frac{H}{\dot{\phi}} d\phi \end{equation} where  $ \phi_{e} $ and $\phi$  are the field values at the end and beginning of Inflation, respectively. Once again using \eqref{24}, we write N as \begin{equation}
	N = \int_{\phi_{e}}^{\phi} \frac{8 \pi}{ m_{p}^{2}} \frac{H^{2} (1 + Q)}{F_{_{\prime} H} H^{\prime}} d\phi
	\end{equation} Now, by the definition , we have \begin{equation}
	dN = \frac{da}{a}
	\end{equation} which gives us \begin{equation}
	a(\phi) = a(\phi_{o}) \exp \left( - \int_{\phi_{e}}^{\phi} \frac{8 \pi}{ m_{p}^{2}} \frac{H^{2} (1 + Q)}{F_{_{\prime} H} H^{\prime}} d\phi \right)
	\end{equation}
	Another important quantity for inflationary models are the slow roll parameters and from the basic definition of the $\epsilon$ slow roll parameter, one can write  \begin{equation}
	\frac{\dddot{a}}{a} = H^{2} + \dot{H} = H^{2} \left( 1 - \epsilon \right) 
	\end{equation} where $ \epsilon $ is the "first" Hubble slow roll parameter given by \begin{equation}
	\epsilon = \frac{m_{p^{2}}}{8 \pi} \frac{F_{_{\prime}H}}{H (1 + Q)}  \left( \frac{H^{\prime}}{H} \right)^{2} 
	\end{equation} This can be arrived at directly by first realizing that \begin{equation} \label{40}
	\dot{H} = H^{\prime} \dot{\phi} = - \frac{m_{p}^{2}}{8 \pi} \frac{F_{_{\prime} H} H^{\prime} }{H (1 + Q)} 
	\end{equation}
	This implies \begin{equation} \label{41}
	-\frac{\dot{H}}{H^{2}} = \frac{m_{p^{2}}}{8 \pi} \frac{F_{_{\prime}H}}{H (1 + Q)}  \left( \frac{H^{\prime}}{H} \right)^{2} 
	\end{equation}
	\\
	$ \epsilon $ is also realized using the concurrent definition \citep{martin2014encyclopaedia} \begin{equation}
	\epsilon = - \frac{d \ln H}{d \ln a} =  \frac{m_{p^{2}}}{8 \pi} \frac{F_{_{\prime}H}}{H (1 + Q)}  \left( \frac{H^{\prime}}{H} \right)^{2} 
	\end{equation}
	The second Hubble Slow Roll Parameter $ \eta $ is similarly defined to be \begin{equation}
	\eta = - \frac{d \ln H^{\prime}}{d \ln a} = \frac{m_{p^{2}}}{8 \pi} \frac{F_{_{\prime}H}}{H (1 + Q)} \frac{H ^{\prime \prime}}{H} 	
	\end{equation}
	\\
	 We now take a brief detour and discuss a bit of subtlety surrounding the definition of slow roll parameters for Warm Inflation. While the $\epsilon$ and $\eta$ slow roll parameters above is derived as it comes from the basic definition \eqref{40} and \eqref{41}, there still exists conflicting views in literature about the appropriate definition for the parameter in the context of Warm Inflation. While some authors \citep{hall2004scalar,visinelli2011natural} like to define the $\epsilon$ and $\eta$ parameters in their usual supercooled inflation forms , others \citep{del2007warm,sayar2017hamilton} like to " absorb" the dependence of the damping function Q as it comes into the slow roll parameters from the basic definitions \eqref{40} and \eqref{41} . The proponents of the former approach feel that defining the first and second slow roll parameters in the usual cold inflation form allows them to have a more relaxed constraint in warm inflationary scenarios \citep{hall2004scalar} . The author of this paper is also in harmony with this line of thought, but however feels the latter approach of defining the first two Hubble parameters in warm inflation allows for them to be evidently more general then their supercooled inflation form. This allows the reader to more smoothly see the transition of the parameters from warm to cold inflation in the extremely low dissipation regime. But we are certainly of the opinion that while both approaches of defining the parameters may appear different on the level of substance, they carry virtually the same essence overall. 
	  So now we just note for the sake of completeness that we can equivalently define the $\epsilon$ and $\eta$ parameters suitable for the approach \citep{hall2004scalar,visinelli2011natural}  as (which we will call $ \epsilon_{a}$ and $ \eta_{a} $) \begin{equation} \label{42}
	  \epsilon_{a} = \frac{m_{p^{2}}}{8 \pi} \frac{F_{_{\prime}H}}{H}  \left( \frac{H^{\prime}}{H} \right)^{2} 
	  \end{equation}
	  \begin{equation} \label{43}
	  	\eta_{a} = \frac{m_{p^{2}}}{8 \pi} \frac{F_{_{\prime}H}}{H} \frac{H ^{\prime \prime}}{H} 
	  \end{equation}
	  which is just the form of the parameters as shown in \citep{del2012approach} . The second form of the parameters , which is in line with the approach of \citep{del2007warm,sayar2017hamilton},  are just the $\eta$ and $\epsilon$ parameters defined in \eqref{40} and \eqref{41}. We readily see that (42) and (43) are just \eqref{40} and \eqref{41} respectively, in the $ Q << 1 $ approximation. 
	Alongside the usual slow roll parameters $\eta$ and $\epsilon$ , \citep{visinelli2011natural} showed that for appropriately studying warm inflationary paradigms some other parameters with the derivatives of $\Gamma$ would be very useful. These parameters were derived by the using the slow roll conditions for warm inflation. Hence we would now like to define some new slow roll parameters for our model, in order to better cater to the needs of warm inflation scenarios. One of the primary slow roll conditions reads \begin{equation} \label{46}
	\frac{- \ddot{H}}{\dot{H} H} << 1
	\end{equation}
	\\
	We would now evaluate the quantity $ \frac{- \ddot{H}}{\dot{H} H} $ to arrive at our new slow roll parameters taking lead from \citep{sayar2017hamilton} . We begin by noting that \begin{equation}
	 \ddot{H} = \frac{d}{dt} \left( H^{\prime} \dot{\phi} \right)
	\end{equation} which can be realized using (38) . Now this allows us to write \begin{equation} \label{48}
	 \ddot{H} = \dot{\phi}^{2} H^{\prime \prime} + \ddot{\phi} H^{\prime}
	\end{equation}
	 To evaluate \eqref{46} , we need to first have an expression for $ \ddot{\phi} $ appropriate for our use. This can be done by using \eqref{24} supplemented by the fact that $ \dot{F_{_{\prime} H}} = \dot{H} F_{_{\prime} H H}  $ where $ \left(  F_{_{\prime} H H} = \frac{\partial^{2} F}{{\partial H}^{2}} \right) $ . A little bit of algebra leads us to \begin{multline}
	 \ddot{\phi} = \frac{-3m_{p}^{2}}{8 \pi} \Bigg( \frac{\left( {H^{\prime}}^{2} F_{_{\prime} H H} + F_{_{\prime} H} H^{\prime \prime} \right) (3H + \Gamma)}{\left( 3H + \Gamma \right)^{2}}  \\ - \frac{(3 H^{\prime} + \Gamma^{\prime} ) F_{_{\prime} H } H^{\prime} }{\left( 3H + \Gamma \right)^{2}} \Bigg)
	 \end{multline} This allows us to write \begin{multline}
	 \frac{- \ddot{H}}{\dot{H} H} = -\eta -  \frac{m_{p}^{2}}{8 \pi H^{2}} \Bigg( \frac{\left( {H^{\prime}}^{2} F_{_{\prime} H H} + F_{_{\prime} H} H^{\prime \prime} \right) (1 + Q ) }{\left( 1 + Q \right)^{2}} \\ - \frac{1}{ \left( 1 + Q \right)^{2} }\frac{ \left(3 H^{\prime} + \Gamma^{\prime} \right) F_{_{\prime} H } H^{\prime}}{3 H} \Bigg) 
	 \end{multline}
	And further we write \begin{multline}
	\frac{- \ddot{H}}{\dot{H} H} = 2 \eta +  \frac{m_{p}^{2} {H^{\prime}}^{2}}{8 \pi H^{2}} \frac{F_{_{\prime} H H}}{1 + Q} - \\ \frac{Q}{1 + Q} \left( \frac{m_{p}^{2}}{8 \pi} \frac{F_{_{\prime} H}}{H} \frac{\Gamma^{\prime} H^{\prime}}{\Gamma H} + \frac{3 m_{p}^{2} F_{_{\prime} H} }{8 \pi H} \frac{H^{\prime} }{\Gamma H (1 + Q)} \right) 
	\end{multline}
	this leads us to define $\beta$ parameter \citep{visinelli2011natural,sayar2017hamilton} as , \begin{equation}
	\beta =   \frac{m_{p}^{2}}{8 \pi} \frac{F_{_{\prime} H}}{H} \frac{\Gamma^{\prime} H^{\prime}}{\Gamma H} \frac{1}{1 + Q }
	\end{equation}
    In addition to these slow roll parameters, we define other parameters which will be helpful for our perturbation spectra studies. These are , \begin{equation}
    \chi =  \frac{ m_{p}^{2} F_{_{\prime} H} }{8 \pi H} \frac{H^{\prime} }{\Gamma H (1 + Q)}
    \end{equation}
    and \begin{equation} \label{54}
    \gamma =  \frac{m_{p}^{2} {H^{\prime}}^{2}}{8 \pi H^{2}} \frac{F_{_{\prime} H H}}{1 + Q}
    \end{equation}
    which now allows us to express \eqref{48} as \begin{equation}
    \frac{- \ddot{H}}{\dot{H} H} = 2 \eta + \chi - \frac{Q}{1 + Q} \left( \beta + 3 \gamma \right)
    \end{equation}
    In addition to these we define more parameters which will be helpful for us in our perturbation spectra analysis
    \begin{equation}
    \delta = \frac{m_{p}^{2}}{8 \pi} \left( \frac{F_{_{\prime} H} }{(1 + Q) H} \right)^{2} \frac{\Gamma^{\prime \prime} {H^{\prime}}^{2}}{\Gamma H^{2}}
    \end{equation}
    \begin{equation}
    \sigma = \frac{m_{p}^{2}}{ 8 \pi} \frac{1}{1 + Q} \frac{F_{\prime} H}{H} \frac{H^{\prime \prime \prime}}{H^{\prime}}
    \end{equation} 
    \begin{equation}
    \psi = \frac{m_{p}^{2}}{ 8 \pi} \frac{F_{_{\prime} H H H}}{(1 + Q)} \frac{{H^{\prime}}^{2}}{H}
    \end{equation}
    This completes the proper dynamical outline of the Exact Solution approach for warm inflation. It is quick to check that this approach reduces to the usual cold inflation approach for exact solutions \citep{del2012approach} in the extremely low dissipation regime $ Q << 1 $ . It also further reduces to the Hamilton-Jacobi method for warm inflation for $ F(H) = H^{2} $ \citep{sayar2017hamilton}, and further  to the usual cold inflation Hamilton-Jacobi approach \citep{kinney1997hamilton} in the extremely low dissipation regime for the same $ F(H) $. With the dynamical aspects covered, we shall now explore the perturbation spectra of warm inflation in this approach.
	\section{ Perturbation spectra analysis }
	
	Cosmological density and gravitational wave perturbations in the inflationary scenario arise as quantum fluctuations
	which redshift to long wavelengths due to rapid cosmological expansion during Inflation \citep{baumann2009tasi,hawking1982development,starobinsky1982dynamics,guth1982fluctuations}. In warm inflation, only the density perturbations couple strongly with the thermal background and hence, the scalar density spectra is the one which looks more evidently different from it's usual supercooled inflationary counterpart \citep{taylor2000perturbation} . Tensor perturbations do not couple strongly to the	thermal background and so gravitational waves are only generated by quantum fluctuations, as in standard cold inflation. In addition to the usual adiabatic perturbations in cold inflation, Isocurvature perturbations also are generated in the warm inflationary era due to thermal fluctuations in the radiation field. These perturbations can be characterised by fluctuations in the entropy of the particle species undergoing thermal fluctuations relative to the number density of photons. But in this paper, we will limit our focus to the study of only adiabatic perturbations.
	\\
	\\
	The square of the amplitude of adiabatic perturbations is calculated in a similar way to cold inflation \citep{taylor2000perturbation,berera2009warm,sayar2017hamilton} , \begin{equation}
	P_{s} (k)^2 = \frac{4}{25} \left(\frac{H}{\dot{|\phi | }} \right)^{2} {d \phi}^{2}
	\end{equation}
	where $ {d \phi}^{2} $ for the high dissipative regime ($ Q >> 1 $ )is given by \citep{berera2000warm} , \begin{equation}
	{d\phi}^{2} = \frac{k_{F} T}{2 \pi}
	\end{equation}  where $ k_{F} $ is the so called freeze out number given by, \begin{equation}
	k_{F} = \sqrt{\Gamma H}
	\end{equation} The definition of the freeze out number $ k_{F} $ is not changed by considering a general Friedmann equation of the form \eqref{19}. It is so because the definition of the freeze out number stems primarily from the field equation of $\phi$ \eqref{12} , in particular from the evolution equation of the fluctuations $ \delta \phi(x,t) $ ( where $\phi$ (x,t) = $ \phi_{o} (x,t) + \delta \phi(x,t)  $ ) \citep{taylor2000perturbation,berera2000warm}.This remains unchanged by the consideration of \eqref{19} , takes the usual form with the inclusion of the spatial Laplacian and an additional white noise random force term , \begin{equation}
	\Gamma \frac{d \delta \phi (k,t)}{dt} = - [\left( k^{2} + V^{\prime \prime} (\phi_{o} ) \right)] \delta \phi (k , t) + \zeta (k,t)
	\end{equation} where we have Fourier transformed to the mpmentum space and $ \zeta (k,t) $ is the white noise term. In a similar way to \citep{taylor2000perturbation} , we reach at $ k_{F} = \sqrt{\Gamma H} $. Now using the expression for T \eqref{28}, and the definition of $\epsilon$ (40) , we can write $ P_{s} (k) $ as \begin{equation}
	{P_{s}(k)}^{2} = \frac{2}{25 \pi} \sqrt{\Gamma H} \left(\frac{3 Q}{4 \sigma}\right)^{\frac{1}{4}} \left( \frac{\epsilon H^{2}}{H^{\prime}}\right)^{\frac{1}{2}} \left(\frac{H^{\prime}}{\epsilon H}\right)^{2} 
	\end{equation} 
	which leads us to write \begin{equation}
		P_{s}(k) = \left(\left(\frac{48}{1562500 \sigma \pi^{4}}\right)^{\frac{1}{2}} \Gamma^{\frac{3}{2}} H^{- 
		\frac{3}{2}} \epsilon^{-3} {H^{\prime}}^{3} \right)^{\frac{1}{2}}
	\end{equation} 
      \\
	The scalar spectral index is defined by the well known equation \begin{equation}
	n_{s} - 1  = \frac{d \ln {P_{s}(k)}^{2}}{d \ln k}
	\end{equation} where $ d \ln k $ is given as the negative of the differential of the number of e-folds \begin{equation}
	d \ln k = - dN = \frac{8 \pi}{3 m_{p}^{2}} \frac{H \Gamma }{F_{_{\prime} H} H^{\prime}} d\phi 
	\end{equation}
	For proceeding further, we note that the derivative of the $\epsilon$ parameter is given by \begin{equation}
	\epsilon^{\prime} = \frac{d \epsilon}{d\phi} = \frac{H^{\prime}}{H} \left( \gamma + 2 \eta - 2 \epsilon - \beta \right) 
	\end{equation} This allows us to write \begin{equation}
	\frac{d \epsilon}{d \ln k} = \epsilon ( \gamma + 2 \eta - 2 \epsilon - \beta)
	\end{equation}
	\begin{equation} \label{69}
	 \frac{d \ln {P_{s}(k)}^{2}}{d \ln k} = \frac{3 \beta}{4} - \frac{3 \epsilon}{4} + \frac{3 \eta}{2} -  \frac{3}{2} \frac{\epsilon^{\prime}}{\epsilon} \frac{3 m_{p}^{2}}{8 \pi} \frac{F_{_{\prime} H} H^{\prime}}{H (\Gamma) }
	\end{equation}
	\begin{equation}
	\frac{d \ln {P_{s}(k)}^{2}}{d \ln k} = \frac{9}{4} \beta + \frac{9}{4} \epsilon - \frac{3}{2} \gamma - \frac{3}{2} \eta
	\end{equation}
	This finally allows to us to express the scalar spectral index as \begin{equation}
	n_{s} = 1 + \frac{9}{4} \beta + \frac{9}{4} \epsilon - \frac{3}{2} \gamma - \frac{3}{2} \eta 
	\end{equation} 
	This is the scalar spectral index for the high dissipative regime.
	\\
	For the low dissipative regime ( $ Q << 1 $ ) we have \begin{equation} \label{72}
	d \phi = H T
	\end{equation}
	which  allows us to write the power spectrum for the low dissipative regime as
	\begin{equation}
	{P_{s}^{\ast}(k)}^{2} = \frac{4}{25} \frac{{H^{\prime}}^{2}}{\epsilon^{2}} T^{2}
	\end{equation}
	Using the expression for the temperature derived above \eqref{28} , we have
	 \begin{equation}
	{P_{s}^{\ast}(k)}^{2} = \left(\frac{4}{625 \alpha} \left(H^{\prime}\right)^{2} H^{3} \Gamma \right)^{\frac{1}{2}} 
	\end{equation} 
	Again , the definition of the scalar spectral index in this case is the same as for the high dissipation scenario \begin{equation}
	n_{s}^{\ast} - 1 = \frac{d \ln {P_{s}^{\ast}(k)}^{2}}{d \ln k}
	\end{equation}
	Pursuing a similar analysis as for the previous case, we arrive at \begin{equation}
	n_{s}^{\ast} = 1 + \frac{3}{2} \epsilon + \eta + \frac{\beta}{2} 
	\end{equation}
     \\
   One of the more exciting findings of the observational data from the Planck, WMAP and COBE experiments is that there is a significant running of the scalar spectral index as well. Traditionally it was taken to be negligible but these experimental findings make them a very important observational quantity. Alongside the Running of the scalar spectral index and the index itself, the tensor to scalar ratio is another important quantity of observational relevance. We will now focus more on the high dissipative regime and calculate the running of the scalar spectral index and the tensor-to-scalar ratio in that limit.   We will not calculate the same in the low dissipative regime but one can calculate them in that limit by pursuing a similar method as we do in the following for the high dissipative regime.          
    \\
    \\
    The running of the spectral index is defined by it's usual definition \begin{equation}
    \alpha_{s} = \frac{d n_{s} }{d \ln k}
    \end{equation} 
     In a similar way as we calculated $ \frac{d \epsilon}{d \ln k} $ in (66), we arrive at the following differentials \begin{equation}
     \frac{d \eta}{d \ln k} = \eta \gamma + \sigma \epsilon - \eta \beta - \eta \epsilon
     \end{equation}
     \begin{equation}
     \frac{d \beta}{d \ln k} = \beta \gamma + \delta + \beta \eta - \beta \epsilon - 2 \beta^{2}
     \end{equation}
     \begin{equation} \label{80}
     \frac{d \gamma}{d \ln k} = \psi \epsilon + \gamma \eta - \gamma \beta - \epsilon \gamma
     \end{equation}
     The above expressions allow us to write the running of scalar spectral index using \eqref{69} as , \begin{equation}
      \alpha_{s} = \frac{15}{4} ( \beta \gamma + \eta \beta + \gamma \epsilon + \eta \epsilon) - \frac{9}{2} ( \epsilon \beta + \epsilon^{2} + \beta^{2} + \eta \gamma ) - \frac{3}{2} (  \psi \epsilon ) + \frac{9}{4} \delta 
     \end{equation} 
     Further, the squared tensor perturbation power spectrum amplitude is defined as \citep{liddle1993cold} \begin{equation} \label{82}
     {P_{T} (k)}^{2} = \frac{32}{75 m_{p}^{4}} V(\phi)
     \end{equation}
     During inflation, $ V \approx \frac{3 m_{p}^{2} F}{8 \pi} $ which is clear by \eqref{21}. The definition of the tensor-to-scalar ratio is \begin{equation}
     r = \frac{{P_{T} (k)}^{2}}{{P_{s}(k)}^{2}}
     \end{equation} We see that \eqref{80} now allows us to  write the tensor-to-scalar ratio as \begin{equation}
     r =   \frac{2 F}{m_{p}^{2}} \left( \frac{4 \alpha}{3} \right)^{\frac{1}{4}} \left( \Gamma^{\frac{-3}{2} } {H^{\prime}}^{\frac{-3}{2}} \epsilon^{3} H^{3} \right)^{\frac{1}{2}}
     \end{equation}
     Again we remark that the above formulations of $ r $ and $ \alpha_{s} $ are specifically for the high dissipative regime $ Q >> 1 $ . One can easily formulate the same for the low dissipative regime $ Q << 1 $ using the same procedure we have shown above. 
     \\
     Now, we have completed all the theoretical basis of our approach. We will now apply it on a Tsallis entropy modified universe. We will study aspects of Warm Inflation in this model in the high dissipative regime.
     \section{Warm inflation in Tsallis entropy modified universe}
     A Tsallis entropy modified universe has the Friedmann equation of the form \eqref{6}. This allows us to write $ F(H) $ for this model as \begin{equation}
     F(H) = H^{2 (2 - \kappa)}
     \end{equation}
     Moving forward we would need to ascertain two more quantities for studying warm inflation in this model, which are the Hubble parameter $ H $ and the dissipation coefficient $ \Gamma $ . Both of them will be taken as functions for $ \phi $ . While this statement can be understood in a straightforward way for $ H $ , for $\Gamma$ the answer could have been different. Usually, $\Gamma$ is taken as a function of only $\phi$ or of both $ \phi $ and the radiation temperature $ T $ \citep{sayar2017hamilton,berera2003construction,hall2004constraining,berera2005absence} . This is because  temperature plays a crucial role in the dissipative scenario of warm inflation. However, temperature in general is written in terms of $\phi$ eventually and the dissipation function in cases with $ \Gamma = \Gamma (\phi , T) $ turns into a function of only $\phi$ \citep{del2009warm,moss2006dissipation,bastero2011warming,herrera2013general} . So in our case we will treat both H and $\Gamma$ as $ H = H(\phi) $ and $ \Gamma = \Gamma (\phi) $ . Further, we consider both H and $\Gamma$ to be power law functions of $\phi$ , as \begin{equation}
     H(\phi) = H_{o} \phi^{n}
     \end{equation}
     \begin{equation}
     \Gamma (\phi) = \Gamma_{o} \phi^{m}
     \end{equation}
     where $ H_{o} $ , $ \Gamma_{o} $ are some constants. The powers n and m are left undetermined here as we will use the Planck data to find out which power laws best fit with our concerned model. The reason for choosing power law form for the Hubble parameter  is because they seem to be a good fit with the latest Planck Data \citep{akrami2018planck} , while the consideration of a supersymmetric interaction of the authors  \citep{berera2003construction,berera2005absence} lead to the dissipation coefficient being a linear function of $\phi$. While other authors from several distinct considerations have been led to power law forms of the dissipation coefficient \citep{del2009warm,herrera2013general,sayar2017hamilton}. This suggests to us that power law forms can indeed be very viable and general forms of the dissipation coefficients. Hence, we take the coefficient to be in a power law form of the field variable.
     \\
     With all the preliminaries cleared up, we now move towards concrete calculations. As stated previously, we will focus on warm inflation in the high dissipative regime for this case. To make progress, we would need the form of the field $\phi$ at the time of horizon exit and at the end of inflation. It is straightforward to get the latter by setting $\epsilon = 1 $ in (40), \begin{equation}
     1 = \frac{3 m_{p}^{2} 2 (2 - \kappa) n^{2} H_{o}^{3 - 2 \kappa} }{8 \pi \Gamma_{o}} \phi_{e}^{(3 - 2 \kappa)n - m - 2}
     \end{equation} 
     \begin{equation}
     \frac{3 m_{p}^{2} 2 (2 - \kappa) n^{2} H_{o}^{3 - 2 \kappa} }{8 \pi \Gamma_{o}} = \phi_{e}^{m + 2 - n(3 - 2 \kappa)}
     \end{equation} 
     In order to get the field at the time of horizon exit, we take the help of the number of e-folds. For our model, it is given by  \begin{equation}
     N = \int_{\phi_{e}}^{\phi} \frac{8 \pi}{3 m_{p}^{2}} \frac{H_{o} \Gamma_{o} \phi^{n} \phi^{m} }{2 (2- \kappa) H_{o}^{3 - 2 \kappa} \phi^{n(3- 2 \kappa) n H_{o} \phi^{n - 1 } } }
     \end{equation}
      This leads to
     \begin{multline}
     {3 m_{p}^{2} 2 (2 - \kappa) n H_{o}^{3 - 2 \kappa} }{8 \pi \Gamma_{o}} N = \\ \frac{1}{m + 2 - n(3 - 2 \kappa) } \left( \phi^{m + 2 -n(3-2 \kappa) } - \phi_{e}^{m + 2 -n(3 - 2\kappa)N }   \right) 
     \end{multline}
     This finally allows us to write the field at the time of horizon exit as  \begin{equation} \label{92}
     \phi^{m + 2 - n(3 - 2 \kappa) } = \phi_{e}^{m + 2 -n(3 - 2\kappa) }  \left( 1 + \frac{(m +2 - n(3 - 2\kappa)) N }{n} \right)
     \end{equation}
     Now, using \eqref{24} we can write \begin{equation}
     \frac{d \phi}{dt} = - \frac{3 m_{p}^{2} 2 (2 - \kappa) H_{o}^{3 - 2 \kappa} \phi^{n(3 - 2\kappa) } n H_{o} \phi^{n-1} }{8 \pi \Gamma_{o} \phi^{m}}
     \end{equation}	
     \\
     Integrating from $ t_{o} $ to some t  , we arrive at  \begin{multline} \label{94}
     \phi^{m+2 - n(4-2\kappa) } (t) = \frac{(4-2\kappa)n - m - 2 }{n} \phi_{e}^{m + 2 - n(3-2\kappa)} H_{o} (t-t_{o} ) + \\ \left(  \phi_{e}^{m + 2 - n(3-2\kappa)} \left( 1 + \frac{(m +2 - n(3 - 2\kappa))N }{n} \right) \right)^{g} 
     \end{multline} 
     where $$g = \frac{(m + 2 - n(4 - 2\kappa)) }{m + 2  - n(3-2 \kappa) } $$
     This immediately allows us to write the Hubble parameter and the dissipation coefficient as functions of time \begin{multline}
     H(t) = H_{o}  \bigg[  \frac{(4-2\kappa)n - m - 2 }{n} \phi_{e}^{m + 2 - n(3-2\kappa)} H_{o} (t-t_{o} ) + \\ \left(  \phi_{e}^{m + 2 - n(3-2\kappa)} \left( 1 + \frac{(m +2 - n(3 - 2\kappa))N }{n} \right) \right)^{g} \bigg]^{nw} 
     \end{multline} 
     \begin{multline}
     \Gamma(t) = \Gamma_{o}  \bigg[  \frac{(4-2\kappa)n - m - 2 }{n} \phi_{e}^{m + 2 - n(3-2\kappa)} H_{o} (t-t_{o} ) + \\ \left(  \phi_{e}^{m + 2 - n(3-2\kappa)} \left( 1 + \frac{(m +2 - n(3 - 2\kappa))N }{n} \right) \right)^{g} \bigg]^{mw} 
     \end{multline} 
     where $$w = \frac{1}{(m + 2 - n(4 - 2 \kappa)) } $$
     The inflationary potential corresponding to this cosmology is given by (30) \begin{multline}
     V(\phi) = \frac{3 m_{p}^{2} H_{o}^{4 - 2 \kappa} \phi^{4 - 2 \kappa} }{8 \pi} - \frac{1}{2} \\ \left( \frac{3m_{p}^{2} 2 (2 - \kappa) H_{o}^{4 - 2 \kappa} n \phi^{(3 - 2 \kappa) n - m - 1} }{8 \pi \Gamma_{o} } \right)^{2} 
     \end{multline}
     By the definition of the Hubble parameter we have the scale factor as a function of time as \begin{multline}
     a(t) = a_{o} \exp \Bigg( \frac{n H_{o} \phi_{e}^{m + 2 - n(3-2\kappa)}}{(m + 2 - (3 - 2 \kappa)n)N} \Bigg[ \phi_{e}^{m + 2 - n(3 - 2\kappa)} \\ \left( 1 + \frac{m +2 - n(3 - 2\kappa) }{n} \right) - \bigg[ \frac{(4-2\kappa)n - m - 2 }{n} \\ \phi_{e}^{m + 2 - n(3-2\kappa)}  H_{o} (t-t_{o} ) + \Bigg(  \phi_{e}^{m + 2 - n(3-2\kappa)} \\ \left( 1 + \frac{(m +2 - n(3 - 2\kappa))N }{n} \right) \Bigg)^{g} \bigg]^{g}   \Bigg]  \Bigg)
     \end{multline}
     We can also get a relationship between the radiation energy density and the inflaton energy density, as a function of $\phi$ and consequently of time. Using \eqref{27} in the high dissipation regime we have, \begin{equation}
     \rho_{r} (\phi) = \frac{3m_{p}^{2} \rho_{\phi} }{32 \pi F} \bigg[ \frac{(F_{_{\prime} H}H^{\prime})^{2}}{\Gamma H}\bigg]
     \end{equation}
     Doing a little bit of algebra on this formula, we arrive at  \begin{equation} \label{103}
     \rho_{r} (\phi) = \frac{ m_{p}^{2} (2 - \kappa) \phi_{e}^{m + 2 -n(3 - 2\kappa) } \rho_{\phi} }{16 \pi \phi^{m + 2 -n(3 -2\kappa)}} 
     \end{equation}
     Using \eqref{94} , we can further write the above expression in terms of time , \begin{multline}
     \rho_{r} (t) = \frac{ m_{p}^{2} (2 - \kappa) \phi_{e}^{m + 2 -n(3 - 2\kappa) } \rho_{\phi} }{16 \pi}  \bigg[  \frac{(4-2\kappa)n - m - 2 }{n} \\ \phi_{e}^{m + 2 - n(3-2\kappa)} H_{o} (t-t_{o} ) + \bigg(  \phi_{e}^{m + 2 - n(3-2\kappa)} \bigg(1 + \\  \frac{(m +2 - n(3 - 2\kappa) )N }{n} \bigg) \bigg)^{g} \bigg]^{g} 
     \end{multline}
     We note further that at the time of horizon exit, \eqref{103} becomes \begin{equation}
     \rho_{r} = \frac{  m_{p}^{2} (2 - \kappa) \rho_{\phi} }{16 \pi } \left( \frac{n}{n + (m + 2 - n(3 - 2\kappa) )N } \right) 
     \end{equation}
     The above expression tells us that in the high dissipation regime at the time of horizon exit, only the free parameters $ \kappa $ , $ m $ and $ n $ and the e-fold number N  determines the relationship between $ \rho_{\phi} $ and $ \rho_{r} $. 
     \\
     \\
     Now in order to fully get the details of Warm Inflation in the high dissipation regime in this cosmology, we would like to have appropriate values of the free parameters $ \kappa $ , $ n $ and $ m $ which fit with the observational data [\citep{akrami2018planck} , \citep{aghanim2018planck}]. We would  like to have expressions for important observational quantities like the scalar spectral index and the running of the scalar spectral index at the time of horizon exit. This can be done by evaluating all the relevant slow roll and other cosmological parameters defined previously at the time of horizon exit . Using \eqref{92}, we can evaluate the slow roll parameters at the time of horizon exit. The $\epsilon$ slow roll parameter in particular is \begin{equation}
     \epsilon = \frac{n}{ n + (m + 2 -n(3-2\kappa)) N }
     \end{equation}
     We have emphasized about the $\epsilon$ parameter here because it is possible to express all the other parameters which we have mentioned before in terms of this parameter. Evaluating all the parameters at horizon exit, we get \begin{equation}
     \beta = \frac{m}{n} \epsilon
     \end{equation}
     \begin{equation}
     \eta = \frac{n-1}{n} \epsilon
     \end{equation}
     \begin{equation}
     \gamma = (3 - 2\kappa) \epsilon
     \end{equation}
     \begin{equation}
     \sigma = \frac{(n-1) (n-2)}{n} \epsilon
     \end{equation}
     \begin{equation}
     \delta = \frac{(m) (m-1)}{n^{2}} \epsilon^{2}
     \end{equation}
     \begin{equation}
     \psi = (3- 2\kappa) (2 - 2 \kappa) \epsilon
     \end{equation}
     Using these definitions and \eqref{72}, we arrive at the following expressions for the scalar spectral index, \begin{equation}
     n_{s} = 1 + \frac{3 (4 \kappa n+3 m-n-2)}{4 N (2 (\kappa-1) n+m+2)} 
     \end{equation}
    And the running of the scalar spectral index is given by (82), \begin{equation}
    \alpha_{s} = -\frac{3 (n (2 (\kappa (4 \kappa-11)+8) n+42 \kappa-55)+51)}{4 ((2 \kappa-3) n N+n+5 N)^2}
    \end{equation}
    With that, we have now studied all the analytical aspects of warm inflation in this scenario.To get more insight into the paradigm of warm inflation in this model, we will have to take note of the latest observational data available from the Planck satellite experiment \citep{aghanim2018planck} and see which values of the free parameters in this model most suitably fit the data available on the spectral index and it's running. Another important quantity in this excursion of ours is the e-fold Number. For inflation to solve cosmological problems and contribute in large scale structure formation, the e-fold number can conveniently between around 60 \citep{baumann2009tasi} , so we will henceforth set $ N=60 $.
    \\
    Using \eqref{72} and \eqref{82} , we find that a suitable choice of the free parameters which satisfies the inflationary Requirements is $ (m , n , \kappa) = (3, -5 , 1.4) $ . For these values , \begin{equation}
    n_{s} \approx 0.964912
    \end{equation}
    \begin{equation}
    \alpha_{s} \approx -0.003
    \end{equation}
    Which is in perfect agreement with the Planck 2018 data \citep{akrami2018planck} of $ n_{s} = 0.9649 \pm 0.0042 $ ( $ 68 \% CL $ , Planck $TT,TE,EE  + lowE + lensing$ ) and the negligible running of the spectral index. Putting these values in the required equations derived above would give one full details of Warm Inflation in this cosmology.One can also further use the constraints on the tensor-to-scalar ratio provided by the data to bound the constants $ H_{o} $ and $ \Gamma_{o} $ in a similar way as done in \citep{sayar2017hamilton} , but we do not pursue that here.  
    
    \section{Conclusions }
    In this paper, we have introduced the exact solution approach for Warm onflation. We started off by showing how many modified cosmological scenarios like braneworld cosmologies , modified gravity Cosmologies and various modified entropy cosmologies have a similar form of the Friedmann equation which can be used to consider a generalized Friedmann equation with a general function of the Hubble parameter . We then began the description of our method with a light review of the basics dynamics of warm inflation. After that, we described our approach and showed how various important quantities for warm inflationary regimes like the Hubble parameter, the dissipation coefficient, the e-fold number, the inflaton field function etc. can be derived using this approach. We further explored scalar and tensorial inflationary perturbations in this method and derived important Inflationary Parameters like the scalar spectral index, it's running, the tensor-to-scalar ratio etc.  Finally, we applied this method to study high dissipation warm inflation in a Tsallis modified entropy  universe. We here point to one peculiarity of our model which we have not yet touched upon. In obtaining the equation of the inflaton field we have assumed that the matter, specified by the inflaton
    scalar field, enters into the action Lagrangian in such a way that its variation in a FLRW background metric leads to the Klein-Gordon equation, expressed by \eqref{12}. Therefore our method is only applicable to theories where the background metric alongside the perturbations, are not modified.This means that Horava-Lifshitz theories of gravity \citep{mukohyama2010hovrava} or theories of similar plight are beyond the scope of our approach. 
    \\
    \\
    \section*{Acknowledgements}
    I would like to thank the referee for their insightful comments on the work, which have increased the depth of the work multi folds.

	\bibliographystyle{cas-model2-names}
	\bibliography{cas-refs}

\end{document}